\documentclass[reprint,aps]{revtex4-1}

\usepackage{graphicx}
\usepackage{amsmath}
\usepackage{hyperref}
\usepackage{tabularx}
\usepackage{hyphenat}
\usepackage[capitalise]{cleveref}

\newcommand{\ket}[1]{\left| #1 \right>}
\newcommand{\bra}[1]{\left< #1 \right|}
\newcommand{\ketbra}[1]{\left| #1 \right>\left< #1 \right|}

\newcommand{\expo}[1]{\text{e}^{ #1 }}

\crefname{equation}{Eq.}{Eqs.}
\Crefname{equation}{Equation}{Equations}
\crefname{figure}{Fig.}{Figs.}
\Crefname{figure}{Figure}{Figures}
\crefname{section}{Sect.}{Sects.}
\Crefname{section}{Section}{Sections}

\newcommand{\ha}{\hat{a}}

\newcommand{\had}{\hat{a}^\dagger}
\newcommand{\hb}{\hat{b}}
\newcommand{\hbd}{\hat{b}^\dag}

\newcommand{\hH}{\hat{H}}

\begin{document}

\title{Fast and Unconditional All-Microwave Reset of a
 Superconducting Qubit}

\author{P.~Magnard$^{1}$, P.~Kurpiers$^{1}$, B.~Royer$^2$, T.~Walter$^1$, J.-C.~Besse$^1$, S.~Gasparinetti$^1$, M.~Pechal$^1$, J.~Heinsoo$^1$, S.~Storz$^1$, A.~Blais$^{2,3}$, A.~Wallraff$^1$}
\affiliation{$^1$Department of Physics, ETH Z\"urich, CH-8093, Z\"urich, Switzerland.}
\affiliation{$^2$Institut Quantique and D\'{e}partement de Physique, Universit\'{e} de Sherbrooke, Sherbrooke J1K 2R1 QC, Canada}
\affiliation{$^3$Canadian Institute for Advanced Research, Toronto, ON, Canada}
\date{\today}

\begin{abstract}
Active qubit reset is a key operation in many quantum algorithms, and particularly in error correction codes. Here, we experimentally demonstrate a reset scheme of a three level transmon artificial atom coupled to a large bandwidth resonator. The reset protocol uses a microwave-induced interaction between the $\ket{f,0}$ and $\ket{g,1}$ states of the coupled transmon-resonator system, with $\ket{g}$ and $\ket{f}$ denoting the ground and second excited states of the transmon, and $\ket{0}$ and $\ket{1}$ the photon Fock states of the resonator. We characterize the reset process and demonstrate reinitialization of the transmon-resonator system to its ground state with $0.2\%$ residual excitation in less than $500 \, \rm{ns}$. Our protocol is of practical interest as it has no requirements on the architecture, beyond those for fast and efficient single-shot readout of the transmon, and does not require feedback.
\end{abstract}
\maketitle

The efficient initialization of a set of qubits into their ground state is one of the basic DiVincenzo criteria~\cite{DiVincenzo2000} required for quantum information processing. It is also critical for the implementation of error correction codes~\cite{Schindler2011,Reed2012,Chiaverini2004} where it is needed to reset ancilla qubits on demand to a fiducial state in short time and with high fidelity. For this reason, qubit reset procedures have been implemented for a wide range of physical quantum computation platforms~\cite{Monroe1995b,Jelezko2004,Elzerman2004,Dutt2007,Rogers2014} including superconducting qubits for which we discuss the most common approaches~\cite{Valenzuela2006,Reed2010,Mariantoni2011a,Johnson2012,Riste2012,Riste2012b,Campagne-Ibarcq2013,Geerlings2013,Mariantoni2011a}.

Reset for superconducting qubits is commonly realized using the outcome of a strong projective measurement to either herald the ground state~\cite{Johnson2012} or deterministically prepare it using feedback~\cite{Riste2012,Riste2012b,Campagne-Ibarcq2013}. The achievable single-shot readout fidelity limits the performance of this approach due to measurement-induced state mixing~\cite{Campagne-Ibarcq2013,Boissonneault2009,Slichter2012}. This effect constrains the quantum-non-demolition nature of dispersive readout giving rise to leakage out of the qubit subspace~\cite{Campagne-Ibarcq2013,Sank2016}, which is particularly detrimental to quantum error correction~\cite{Fowler2013}.

Alternatively, qubit reset can be achieved by coupling its excited state to a rapidly decaying quantum system with less thermal excitation. Such driven reset schemes~\cite{Valenzuela2006,Grajcar2008,Reed2010,Geerlings2013} make use of ideas related to dissipation engineering~\cite{Murch2012a,Premaratne2017a,Liu2016c}. In one variant of this approach~\cite{Reed2010}, qubits are quickly tuned into resonance with a Purcell filtered, large-bandwidth, resonator using magnetic flux.  The qubit coupled to the resonator quickly thermalizes to its ground state due to Purcell decay, the rate of which can be adjusted, on-demand, by three orders of magnitude. The flux pulses employed in this scheme require careful calibration and may affect subsequent gates by bleedthrough and neighboring qubits through cross-talk~\cite{Kelly2014}.

An all-microwave reset protocol utilizing the qubit-state-dependent response of a resonator~\cite{Geerlings2013} avoids the use of flux tuning and its potentially detrimental side effects. This protocol has minimal hardware requirements, only a single resonator, but requires a cavity linewidth $\kappa$ smaller than the dispersive interaction strength  $\chi$ limiting both the speed of the reset process and the readout in case the same resontor is used~\cite{Gambetta2007,Walter2017}.

In this work, we demonstrate an alternative all-microwave reset protocol of a three-level transmon coupled to a resonator with no constraint on $\kappa$. Driving the transmon simultaneously with two coherent tones forms a $\Lambda$ system in the Jaynes-Cumming ladder~\cite{Pechal2014} that unconditionally transfers any excitation in the three lowest energy levels of the transmon to a single-photon emitted to the environment, thus resetting the transmon qutrit on-demand. We implement this protocol in an architecture we designed for rapid and high-fidelity transmon readout~\cite{Walter2017}. In addition, the implemented protocol outperforms existing measurement-based and all-microwave driven reset schemes in speed and fidelity (\cref{app:overview}), populates the resonator with one photon at most, and can be extended to other types of superconducting qubits. 

\begin{figure}[t]
\centering
\includegraphics{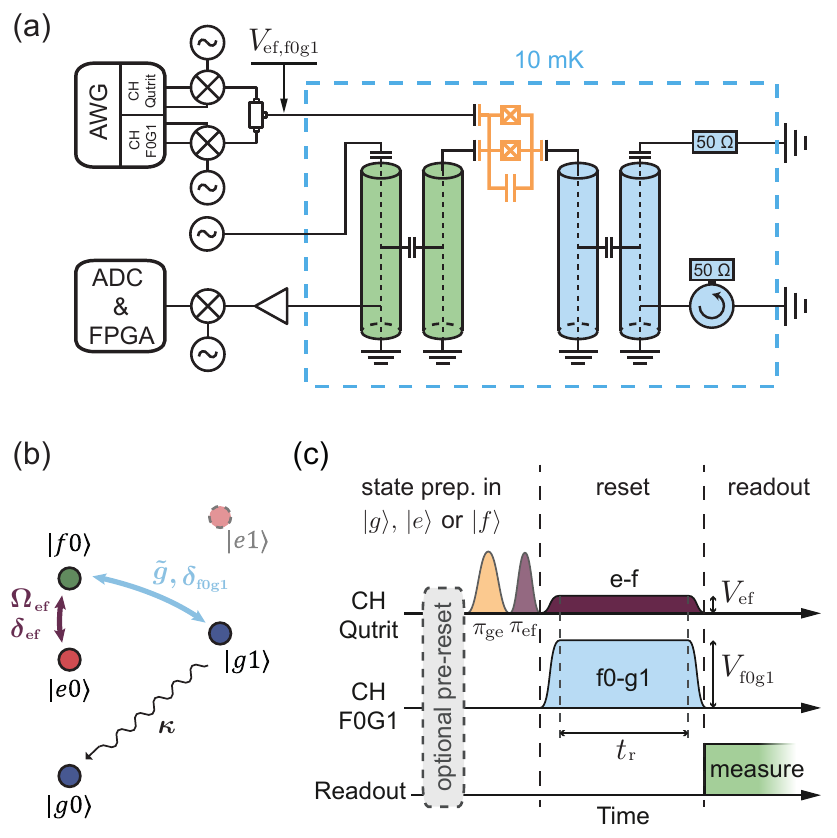}
\caption{ 
(a) Simplified schematic of the experimental setup. A transmon (orange) is coupled to two Purcell filtered resonators. The readout resonator (green) is connected to room temperature electronics (description in the main text), while the reset resonator (blue) is connected to two \(50\,\Omega\) ports thermalized at base temperature. 
(b) Jaynes-Cummings ladder diagram of the transmon/reset resonator energy levels. The purple and light blue arrows represent the e-f and f0-g1 pulsed coherent drives, respectively, and the black arrow labelled \(\kappa\)illustrates the resonator decay process.
(c) Illustration of the pulse schemes used to test the reset protocol. The qutrit is initialized to its ground state passively or optionally with an unconditional reset, then prepared in the desired state $\ket{g}$, $\ket{e}$ or $\ket{f}$ with control pulses (labeled \(\pi_\mathrm{ge}\) and \(\pi_\mathrm{ef}\)). The qutrit is reset by simultaneously applying square e-f (purple) and f0-g1 (light blue) pulses for a reset time \(t_{\rm{r}}\). The resulting qutrit state is then measured by applying a microwave tone to the readout resonator (green). 
}
\label{fig:diagrams}
\end{figure}

The device used in our experiment and schematically illustrated in \cref{fig:diagrams}~a, uses a transmon qubit~\cite{Koch2007,Schreier2008} (orange), with transition frequency $\omega_\mathrm{ge}/2\pi = 6.343\,\mathrm{GHz}$, anharmonicity $\alpha/2\pi = -265\,\mathrm{MHz}$, energy relaxation times $T_1^\mathrm{ge} = 5.5\,\mu\mathrm{s}$ and $T_1^\mathrm{ef} = 2.1\,\mu\mathrm{s}$, and coherence times $T_2^\mathrm{ge} = 7.6\,\mu\mathrm{s}$ and $T_2^\mathrm{ef} = 4.2\,\mu\mathrm{s}$ (\cref{app:params}).
The qubit state is controlled with microwave pulses up-converted from an arbitrary waveform generator (AWG), and applied to the transmon through a dedicated drive line. 
To perform the reset, the transmon is capacitively coupled with rate $g_\mathrm{r}/2\pi = 335\,\mathrm{MHz}$ to a Purcell filtered resonator (light blue) of frequency $\omega_\mathrm{r}/2\pi = 8.400 \,\mathrm{GHz}$ and effective coupling $\kappa/2\pi = 9\,\mathrm{MHz}$ to $50\,\Omega$ ports with cold incoming thermal fields.
To perform readout, the transmon is capacitively coupled with rate $g_\mathrm{m}/2\pi = 210\,\mathrm{MHz}$ to a dedicated, Purcell filtered, resonator (light green) of frequency $\omega_\mathrm{m}/2\pi = 4.787 \,\mathrm{GHz}$ and effective coupling $\kappa_\mathrm{m}/2\pi = 12.6\,\mathrm{MHz}$ to the measurement output line.
The detuning $\Delta_\mathrm{r/m}=\omega_\mathrm{ge}-\omega_\mathrm{r/m}$ between the transmon and both the reset and readout resonators being larger than the coupling rates $g_\mathrm{r/m}$ results in dispersive interactions characterized by the rates $\chi_\mathrm{r}/2\pi = -6.3\,\mathrm{MHz}$ and $\chi_\mathrm{m}/2\pi = -5.8\,\mathrm{MHz}$, respectively.

We read out the transmon state using a gated drive applied to the input port of the readout resonator with a frequency $\omega_\mathrm{d}/2\pi = 4.778\,\mathrm{GHz}$ optimized for qutrit readout~\cite{Bianchetti2010}. The signal scattered off the readout resonator is amplified at $T_{\rm{BT}}=10\,\mathrm{mK}$ by a Josephson parametric amplifier (JPA)~\cite{Yurke1996,Eichler2014} with $20\,\mathrm{dB}$ gain, $20\,\mathrm{MHz}$ bandwidth and a phase-preserving detection efficiency $\eta = 0.61$. We cancel the reflected pump tone interferometrically. The signal is then band-pass filtered and amplified at $4\,\mathrm{K}$ with a high electron mobility transistor (HEMT), down-converted using an I-Q mixer, digitized using an analog-to-digital converter (ADC), digitally down-converted and processed using a field programmable gate array (FPGA).

The reset concept, illustrated in \cref{fig:diagrams}b, is based on a cavity-assisted Raman transition between $\ket{f,0}$ and $\ket{g,1}$~\cite{Pechal2014,Zeytinoglu2015,Gasparinetti2016}. Here $\ket{s,n}$ denotes the tensor product of the transmon in state $\ket{s}$, with $\ket{g}$, $\ket{e}$ and $\ket{f}$ its three lowest energy eigenstates, and the reset resonator in the $n$ photon Fock state $\ket{n}$. By simultaneously driving the $\ket{f,0} \leftrightarrow \ket{g,1}$ (f0-g1) transition and the $\ket{e,0} \leftrightarrow \ket{f,0}$ (e-f) transition, the population is transferred from the qutrit excited states, $\ket{e,0}$ and $\ket{f,0}$, to the state $\ket{g,1}$. The system then rapidly decays to the target dark state $\ket{g,0}$ by photon emission at rate~$\kappa$, effectively resetting the qubit to its ground state.
\begin{figure}[h!]
\centering
\includegraphics{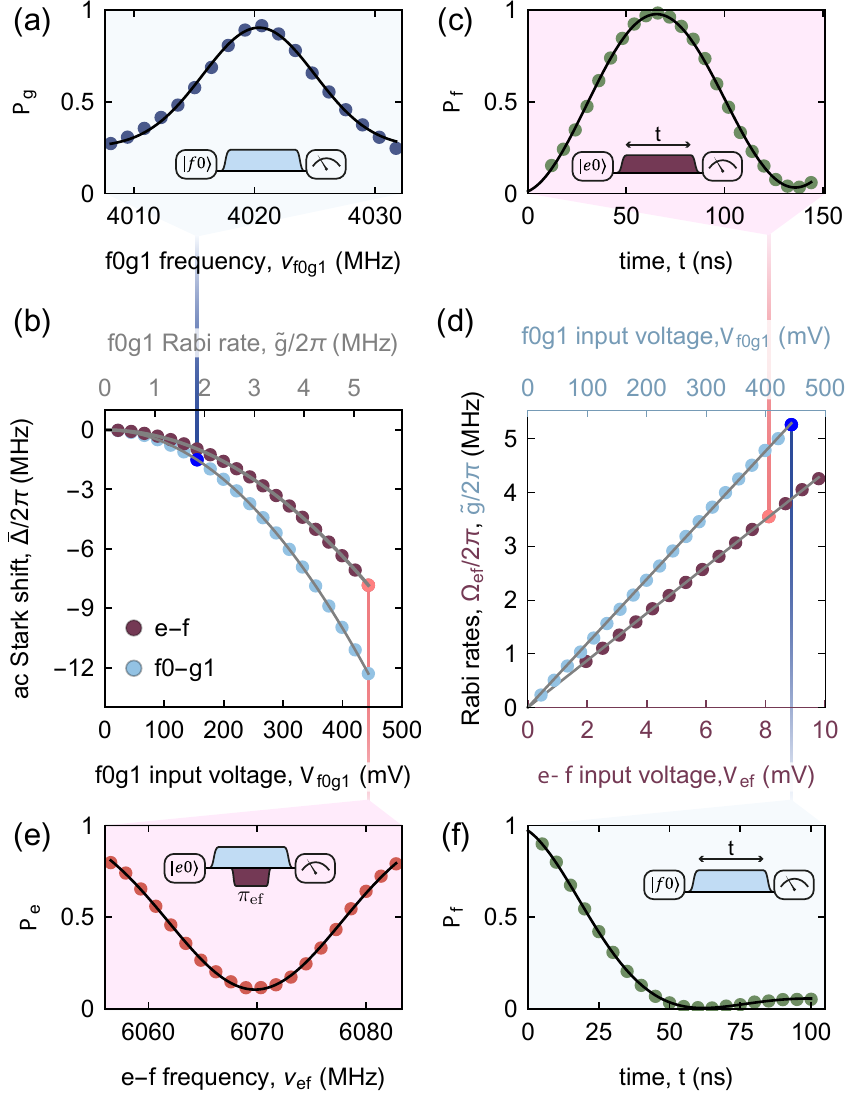}
\caption{
(a) Population \(P_\mathrm{g}\)~\textsl{vs.}~the frequency \(\nu_\mathrm{f0g1}\) of a \(171\,\mathrm{ns}\) long square f0-g1 pulse, of input voltage amplitude \(V_\mathrm{f0g1} = 155\,\mathrm{mV}\).
(b) Measured ac Stark shifts \(\bar{\Delta}_\mathrm{f0g1}\) and \(\bar{\Delta}_\mathrm{ef}\)of the f0-g1 (light blue) and e-f (purple) transitions, \textsl{vs.}~amplitude \(V_\mathrm{f0g1}\) of the f0-g1 drive. The solid lines are quadratic fits to the data.
(c) Population \(P_\mathrm{f}\) \textsl{vs.} the duration \(t\) of a resonant square e-f pulse, of amplitude \(V_\mathrm{ef} = 8\,\mathrm{mV}\), applied to the qutrit initially prepared in $\ket{e,0}$. 
(d) Extracted Rabi rates \(\Omega_\mathrm{ef}\) and \(\tilde{g}\), of the e-f (purple) and f0-g1 (light blue) drives versus their amplitude, \(V_\mathrm{ef}\) and \(V_\mathrm{f0g1}\). The solid lines are linear fits.
(e) Population \(P_\mathrm{e}\) \textsl{vs.}~frequency \(\nu_\mathrm{ef}\) of a square e-f \(\pi\)-pulse applied on the qutrit, initially prepared in state  $\ket{e,0}$, in the presence of a continuous f0-g1 drive of amplitude \(V_\mathrm{f0g1} = 444\,\mathrm{mV}\). 
(f) Population \(P_\mathrm{f}\) \textsl{vs.} the duration \(t\) of a resonant square f0-g1 pulse, of amplitude \(V_\mathrm{f0g1} = 444\,\mathrm{mV}\), applied to the qutrit initially prepared in  $\ket{f,0}$.
The pulse scheme used to acquire the data shown in panels (a), (b), (c) and (d) are shown as insets, with the f0-g1 and e-f pulse envelopes represented in light blue and purple, respectively.
The solid lines in (a) and (e) are fits to a Gaussian from which we extract the resonant frequency of of the e-f and f0-g1 transitions.
The solid lines in (c) and (f) are fits to Rabi oscillations, based on models described in \cref{app:fits}, from which the Rabi rates shown in (d) are extracted.
}
\label{fig:resetCalib}
\end{figure}

We developed a 4 step calibration procedure to accurately determine ac Stark shifts dependent on the f0-g1 drive amplitude~\cite{Zeytinoglu2015}, and the relation between the amplitude and the Rabi rate of each drive. This calibration enables full control over the four tunable parameters which determine the reset dynamics: the detunings $\delta_\mathrm{ef}$ and $\delta_\mathrm{f0g1}$ of the two drive tones to the e-f and f0-g1 ac Stark shifted transition, and their Rabi rates $\Omega_\mathrm{ef}$ and $\tilde{g}$ (\cref{fig:diagrams}b).
In all calibration measurements, the populations $P_\mathrm{g,e,f}$ of the transmon qutrit are extracted by comparing the averaged signal transmitted through the readout resonator to reference traces~\cite{Bianchetti2010}.

As a first step, we determine the ac Stark shift of the f0-g1 transition induced by the f0-g1 drive. We initialize the transmon in $\ket{g}$, then apply a sequence of two $\pi$-pulses ($\pi_\mathrm{ge}$, $\pi_\mathrm{ef}$) to prepare the system in $|f,0\rangle$  (\cref{fig:diagrams}c). 
We apply a flat top f0-g1 pulse with Gaussian rising and falling edges, of carrier frequency $\nu_\mathrm{f0g1}$, amplitude $V_\mathrm{f0g1}$ and duration $t_\mathrm{r}$ and read out the resulting transmon state. 
We repeat the process varying $V_\mathrm{f0g1}$ and $\nu_\mathrm{f0g1}$, and keeping $V_\mathrm{f0g1} t_\mathrm{r}$ fixed to obtain comparable Rabi angle of the rotations induced by the f0-g1 drive. 
For a given value of $V_\mathrm{f0g1}$, we fit the dependence of $P_\mathrm{g}$ on $\nu_\mathrm{f0g1}$ to a Gaussian whose center, at which the population transfer from $|f,0\rangle$ to $|g,1\rangle$ is maximal, yields the ac Stark shifted frequency (\cref{fig:resetCalib}a). 
The ac Stark shift $\bar{\Delta}_\mathrm{f0g1}$ extracted in this way shows a quadratic dependence on $V_\mathrm{f0g1}$ (light blue dots in \cref{fig:resetCalib}b).

In a second step, we determine the Rabi rate $\Omega_\mathrm{ef}$, by preparing the system in $|e,0\rangle$, applying a square, resonant e-f pulse of amplitude $V_\mathrm{ef}$ and duration $t$, and reading out the transmon. For each $V_\mathrm{ef}$, the extracted $\ket{f}$ population $P_\mathrm{f}$ oscillates as a function of $t$ at the Rabi rate $\Omega_\mathrm{ef}$ (\cref{fig:resetCalib}c).
We extract $\Omega_\mathrm{ef}$ by fitting the data to a Rabi model (\cref{app:fits}), and find that it scales linearly in $V_\mathrm{ef}$ (purple dots in \cref{fig:resetCalib}d).

In a third step, we calibrate the ac Stark shift $\bar{\Delta}_\mathrm{ef}$ of the e-f transition induced by the f0-g1 drive. We prepare the system in $|e,0\rangle$, then drive the qutrit with a resonant, square f0-g1 pulse with amplitude $V_\mathrm{f0g1}$ for $420\,\mathrm{ns}$, in the middle of which we simultaneously apply a $140\,\mathrm{ns}$ long square e-f $\pi$-pulse of frequency $\nu_\mathrm{ef}$. 
As before, for each $V_\mathrm{f0g1}$, we extract the ac Stark shifted frequency of the e-f transition by fitting $P_\mathrm{e}$ \textsl{vs.}~$\nu_\mathrm{ef}$ to a Gaussian (\cref{fig:resetCalib}e). $\bar{\Delta}_\mathrm{ef}$ also shows a quadratic dependence on $V_\mathrm{f0g1}$ (purple dots in \cref{fig:resetCalib}b).

In the fourth and final step, we determine the dependence of $\tilde{g}$ on $V_\mathrm{f0g1}$.
The system is initialized in $|f,0\rangle$, then driven with a square, resonant, f0-g1 pulse of amplitude $V_\mathrm{f0g1}$ and duration $t$. The population $P_\mathrm{f}$, extracted at the end of the pulse sequence, exhibits damped oscillations as a function of $t$ (\cref{fig:resetCalib}f), which results from the Rabi oscillation induced between $|f,0\rangle$ and $|g,1\rangle$ at rate $\tilde{g}$ in the presence of spontaneous decay from $|g,1\rangle$ to $|g,0\rangle$ at rate $\kappa$. We fit a Rabi model with loss to $P_\mathrm{f}$~\textsl{vs.}~$t$ to extract $\tilde{g}$ for each $V_\mathrm{f0g1}$, as well as $\kappa$ (\cref{app:fits}), and find a linear dependence of $\tilde{g}$ on $V_\mathrm{f0g1}$ (light blue dots in \cref{fig:resetCalib}d).

\begin{figure}[t!]
\includegraphics{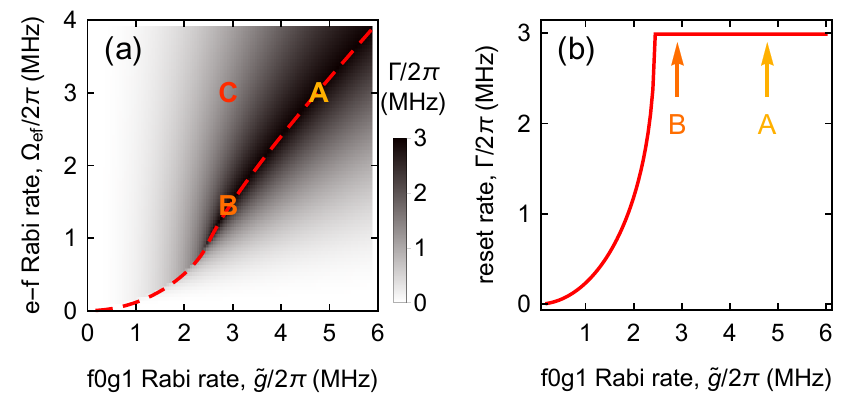}
\caption{(a) Calculated reset rate \(\Gamma\), \textsl{vs.}~Rabi rates \(\tilde{g}\) and \(\Omega_\mathrm{ef}\). The red dashed line shows the value of \(\Omega_\mathrm{ef}\) maximizing the reset rate \(\Gamma\) as a function of \(\tilde{g}\).
(b) Reset rate \(\Gamma\) \textsl{vs.}~\(\tilde{g}\), for an optimal choice of \(\Omega_\mathrm{ef}\) (we follow the red line from (a)).
The parameter configurations A, B and C at which the reset dynamic was probed (see main text and \cref{fig:results}) are indicated with colored letters in (a) and (b).}
\label{fig:resetRate}
\end{figure}

\begin{figure}[t!]
\centering
\includegraphics{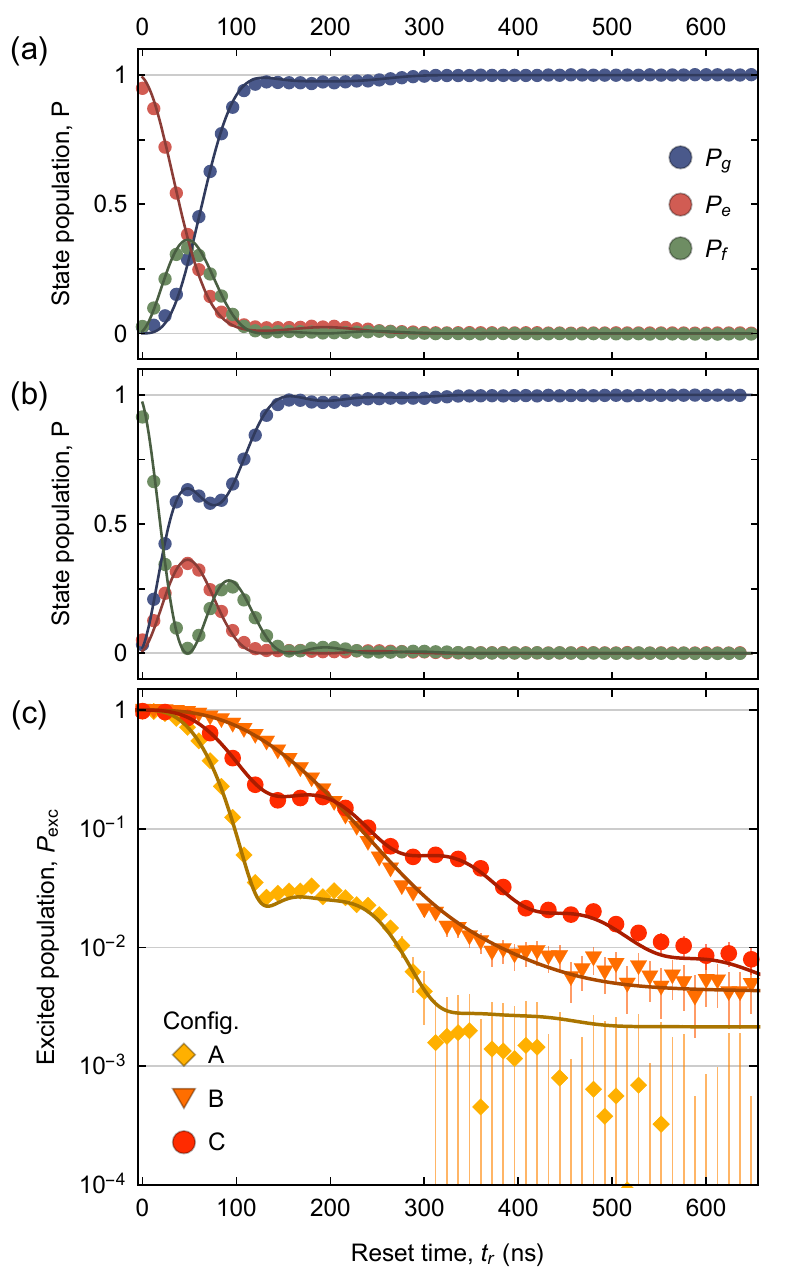}
\caption{(a) Qutrit populations \(P_\mathrm{g,e,f}\) \textsl{vs.} reset time \(t_\mathrm{r}\) with reset parameters in configuration A (see main text), with the system initialized in  $\ket{e,0}$. (b) Same as (a) with the system initialized in $\ket{f,0}$. The solid lines in (a) and (b) are calculated from Hamiltonian~\ref{eq:ResetDynamics}, using the parameters extracted with the Rabi rate calibrations (\cref{fig:resetCalib}d). (c) Excited population \(P_\mathrm{exc}\) as a function of reset time \(t_\mathrm{r}\), when the qutrit is initialized in  $\ket{e}$, shown for reset parameter configurations A, B and C. The solid lines are results of a master equation simulation.}
\label{fig:results}
\end{figure}

In all following experiments, we calibrate all drive frequencies to enforce $\delta_\mathrm{ef} = \delta_\mathrm{f0g1}=0$ to reset the full three-level transmon, leaving only $\tilde{g}$ and $\Omega_\mathrm{ef}$ as tunable parameters.
We determine those Rabi rates that optimize the unconditional reset protocol 
by modelling this process with the non-Hermitian Hamiltonian	
\begin{equation}
	H = 
	  \begin{bmatrix}
    0 										& \Omega_\mathrm{ef} & 0  			 \\
    \Omega_\mathrm{ef}^*  & 0 								 & \tilde{g} \\
		0											& \tilde{g}^*				 & i \kappa/2
  \end{bmatrix}
\label{eq:ResetDynamics}
\end{equation}
acting on the states $\ket{e,0}$, $\ket{f,0}$ and $\ket{g,1}$. The imaginary part of the eigenvalues of this Hamiltonian $\rm{Im}(\lambda_\mathrm{i})$ correspond to half the decay rates to the dark state $\ket{g,0}$. It is therefore natural to define the reset rate $\Gamma\equiv 2\min |\rm{Im}(\lambda_\mathrm{i})|$ as the smallest of these decay rates. 
Plotting $\Gamma$ as a function of $\tilde{g}$ and $\Omega_\mathrm{ef}$ (\cref{fig:resetRate}a), we notice that for each $\tilde{g}$ we find a unique $\Omega_\mathrm{ef}=\Omega_\mathrm{ef}^\mathrm{opt}(\tilde{g})$ that maximizes $\Gamma$, shown as a red dashed line in \cref{fig:resetRate}a. Following this line, $\Gamma$ first increases with $\tilde{g}$ and abruptly reaches its maximal values $\kappa/3 = 2\pi\times 3\,\mathrm{MHz}$, showing a plateau where the reset rate is limited by $\kappa$ (\cref{fig:resetRate}b).

We probed the reset dynamics for two parameter configurations, labelled A and B, which maximize $\Gamma$ and correspond to drive rates $\{\Omega_\mathrm{ef},\tilde{g}\}/2\pi$ set to $\{3,4.8\}$ and $\{1.5,2.9\}\,\mathrm{MHz}$, respectively (\cref{fig:resetRate}a).
We also probed a configuration, labelled C, with Rabi rates $\{3,2.9\}\,\mathrm{MHz}$ that does not maximize $\Gamma$. Configuration differs from B by a higher rate $\Omega_\mathrm{ef}>\Omega_\mathrm{ef}^\mathrm{opt}(\tilde{g})$ (\cref{fig:resetRate}a).
As illustrated in \cref{fig:diagrams}c, we initialize the transmon in $\ket{e,0}$ or $\ket{f,0}$, apply the reset drive pulses for a time $t_\mathrm{r}$, and then readout the transmon with single-shot measurements. Utilizing the single-shot statistics, we can correct for the qutrit state assignment errors, to determine the population of the qutrit with systematic errors below $0.3\%$ (\cref{app:readout}). 
As illustrated in \cref{fig:results}a and b for configuration A, the transmon state oscillates between $\ket{g}$, $\ket{e}$ and $\ket{f}$ while rapidly decaying to $\ket{g}$ on a scale of $100\,\mathrm{ns}$ independent of the initial state. 
The reset dynamics calculated from Hamiltonian~(\ref{eq:ResetDynamics}) is in excellent agreement with the data, as shown by the solid lines in \cref{fig:results}a and b. 
For the most efficient parameter configuration (A), the excited-state population $P_\mathrm{exc}=P_{\rm{e}}+P_{\rm{f}}$ drops below $1\%$ in only $280\,\mathrm{ns}$, and below measurement errors ($\sim 0.3\%$) at steady-state (\cref{fig:results}c), outperforming all existing measurement-based and microwave-driven reset schemes by an order of magnitude (\cref{app:overview}).
Configuration B has smaller drive rates than configuration A. Therefore, the decaying state $\ket{g,1}$ starts being populated at a later time, which explains why $P_\mathrm{exc}$ drops at a later time than configuration A, despite reaching the same maximal reset rate $\Gamma = \kappa/3$. Configuration B is close to a parameter regime where the eigenvalues of Hamiltonian~(\ref{eq:ResetDynamics}) become purely imaginary, which explains why it leads to a reset with almost no oscillatory features.
Configuration C has a higher value of $\Omega_\mathrm{ef}$ than configuration B, but yields a smaller $\Gamma$. This leads $P_\mathrm{exc}$ to drop earlier but at slower rate. 
At long reset times, $P_\mathrm{exc}$ saturates to a non-zero steady-state value $P_\mathrm{exc}^\mathrm{sat}$ because of transmon thermalization, residual driving of the g-e transition by the e-f drive, and finite resonator temperature (\cref{app:limits}). 
To fully capture the role of decoherence and thermalization during the reset, we perform a master equation simulation of the process using only parameters extracted from independent measurements (\cref{app:MES}).
The numerical simulations, shown as solid lines in \cref{fig:results}c, yield $P_{\rm{exc}}^{\rm{sat}}=0.2\%$ for configuration A, and suggest that the achievable $P_{\rm{exc}}^{\rm{sat}}$ is limited by transmon thermalization for all parameter configurations. 
The excellent agreement between the simulation and the data, for all probed reset parameter configurations demonstrates our high level of control and understanding of the process.

In this experiment, we decoupled the readout from the reset process by using two independent sets of resonators and Purcell filters. In practice, the readout resonator can be used to reset the transmon. In this way, the presented unconditional reset protocol 
would benefit from optimizing the design parameters for transmon readout.  
Specifically, high transmon anharmonicity combined with large transmon-resonator coupling $g$ enables one to reach larger $\tilde{g}$ with lower drive amplitude~\cite{Zeytinoglu2015}. Further, a larger resonator bandwidth $\kappa$ leads to faster and higher fidelity reset. We have shown that increasing these parameters also optimizes speed and fidelity of qubit readout~\cite{Walter2017}. 
Using the results of the work presented here, we calculate that implementing this reset protocol with the readout resonator of Ref.~\cite{Walter2017} would lead to the high reset rate $\Gamma = \kappa/3 = 2\pi \times 12.5\,\rm{MHz}$. The qutrit excited population $P_{\rm{exc}}$ would reach $0.1\%$ in $83\,\rm{ns}$, and a steady-state value $P_{\rm{exc}}^{\rm{sat}} = 1.6\times10^{-4}$ in $200\,\rm{ns}$.

In conclusion, we have demonstrated an unconditional all-microwave protocol to reset the state of a three-level transmon below $1\%$ excitation in les than $280\,\rm{ns}$.
This reset scheme 
does neither require feedback, qubit tunability, strict constrains between parameters nor populating the readout resonator with a large numbers of photons.
Furthermore, the 
protocol can conveniently be integrated in an architecture where the qubits are coupled to high bandwidth, Purcell filtered resonators, in order to perform rapid and high-fidelity operations~\cite{Kurpiers2017a} and readout~\cite{Walter2017,Heinsoo2018}.  

We thank Christian Kraglund Andersen for helpful discussions. This work is supported by the European Research Council (ERC) through the "Superconducting Quantum Networks" (SuperQuNet) project, by National Centre of Competence in Research "Quantum Science and Technology" (NCCR QSIT), a research instrument of the Swiss National Science Foundation (SNSF), by the Office of the Director of National Intelligence (ODNI), Intelligence Advanced Research Projects Activity (IARPA), via the U.S. Army Research Office grant W911NF-16-1-0071, NSERC, the Canada First Research Excellence Fund and the Vanier Canada Graduate Scholarships and by ETH Zurich. The views and conclusions contained herein are those of the authors and should not be interpreted as necessarily representing the official policies or endorsements, either expressed or implied, of the ODNI, IARPA, or the U.S. Government. The U.S. Government is authorized to reproduce and distribute reprints for Governmental purposes notwithstanding any copyright annotation thereon.

\appendix

\section{Performance of Reset Protocols for Superconducting Qubits.}
\label{app:overview}
\begin{figure}[b!]
\centering
\includegraphics{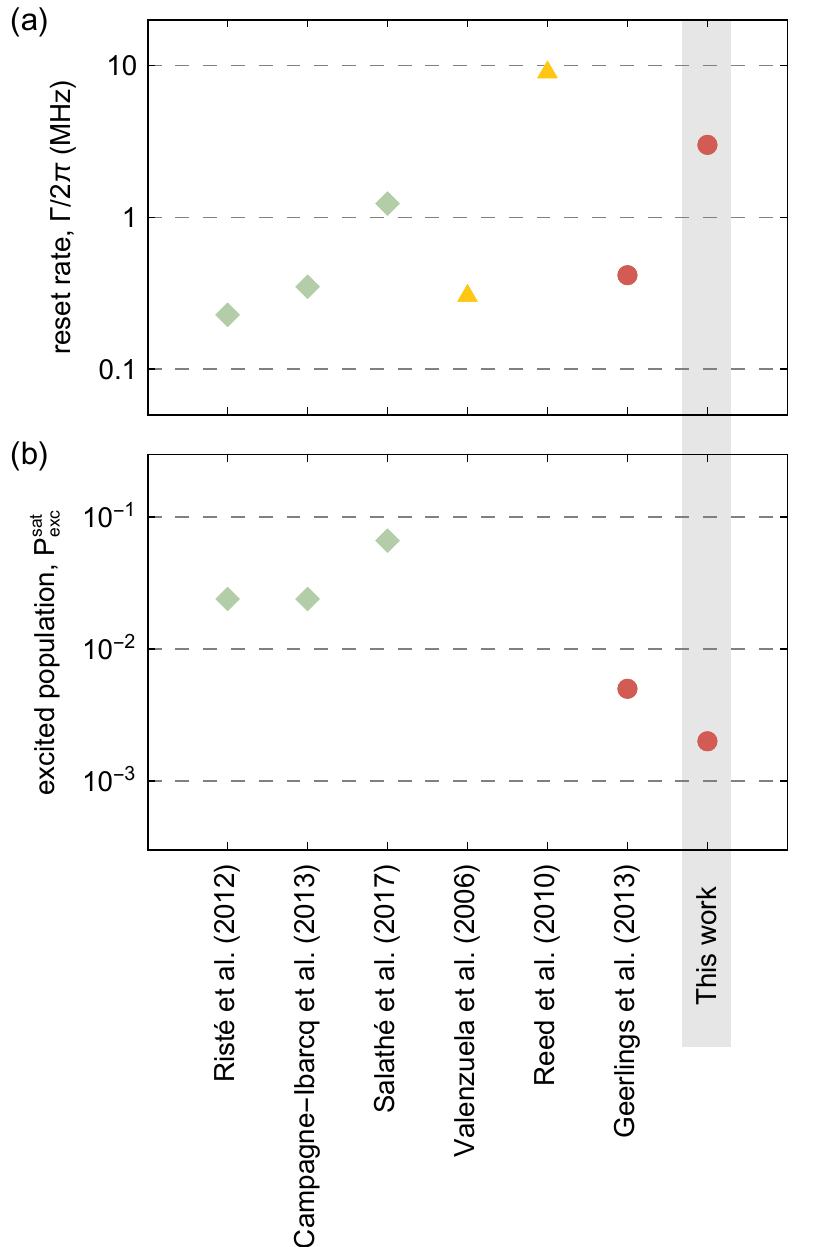}
\caption{
Experimentally achieved reset rates \(\Gamma\) (a) and residual excited state populations \(P_{\rm{exc}}^{\rm{sat}}\) (b) of selected implementations of superconducting qubit reset protocols based on: qubit measurement and feedback control (green squares)~\cite{Riste2012,Campagne-Ibarcq2013,Salathe2017}, qubit frequency tuning via flux pulses (yellow triangles)~\cite{Valenzuela2006,Reed2010} and all-microwave drive induced dissipation (red circles)~\cite{Geerlings2013}. }
\label{fig:overview}
\end{figure}
We compare experimental implementations of superconducting qubit reset protocols by the relevant performance metrics, reset rate \(\Gamma\) and residual excited state population \(P_{\rm{exc}}^{\rm{sat}}\) (\cref{fig:overview}). \(P_{\rm{exc}}^{\rm{sat}}\) is obtained at the end of the reset procedure (measurement-based reset) or at steady-state (driven reset), corresponding in all cases to the lowest residual excitation reached. For driven reset protocols, \(\Gamma\) is defined as the rate at which the qubit approaches the ground state. For measurement based protocols, \(\Gamma\) satisfies \( P_{\rm{exc}}^{\rm{sat}} = e^{-\Gamma t_p}\), where \(t_p\) is the total initialization time from the beginning of the measurement to the end of the conditional control pulse.

\section{Sample Parameters}
\label{app:params}
\begin{table}[b]
\begin{tabular}{|l|r |l|r|}
\hline 
$\omega_{m}/2\pi$      & 4.787~GHz       &$\omega_{r}/2\pi$       & 8.400~GHz      \tabularnewline
$\omega_{PFm}/2\pi$    & 4.778~GHz       &$\omega_{PFr}/2\pi$     & 8.443~GHz      \tabularnewline
$\rm{Q}_{PFr}$         & 91              &$\rm{Q}_{PFr}$          & 60             \tabularnewline
$\rm{J}_{m}/2\pi$      & 13.6~MHz        &$\rm{J}_{r}/2\pi$       & 20.9~MHz       \tabularnewline
$\kappa_{m}/2\pi$      & 12.6~MHz        &$\kappa/2\pi$           & 9.0~MHz        \tabularnewline
$\omega_{\rm{ge}}/2\pi$& 6.343~GHz       &                        &                \tabularnewline
$\alpha/2\pi$          & -265~MHz        &                        &                \tabularnewline 
$n_\mathrm{th}$        & 17~\%           &                        &                \tabularnewline 
$\chi_{m}/2\pi$        & -5.8~MHz        &$\chi_{r}/2\pi$         & -6.3~MHz       \tabularnewline
$\rm{g}_{m}/2\pi$      & 210~MHz         &$\rm{g}_{r}/2\pi$       & 335~MHz        \tabularnewline
$\rm{T}^{\rm{ge}}_{1}$ & 5.5~$\mu\rm{s}$ &$\rm{T}^{\rm{ef}}_{1}$  & 2.1~$\mu\rm{s}$\tabularnewline
$\rm{T}^{\rm{ge}}_{2}$ & 7.6~$\mu\rm{s}$ &$\rm{T}^{\rm{ef}}_{2}$  & 4.2~$\mu\rm{s}$\tabularnewline
$\rm{T}^{*\rm{ge}}_{2}$& 3.5~$\mu\rm{s}$ &$\rm{T}^{*\rm{ef}}_{2}$ & 2.0~$\mu\rm{s}$\tabularnewline
\hline 
\end{tabular}
\caption{Sample parameters: From time resolved Ramsey measurements we extract the ge transition frequency $\omega_{\rm{ge}}/2\pi$, and the anharmonicity $\alpha/2\pi$. 
From resonator transmission spectroscopy we obtain the frequencies, quality factors and couplings of the measurement (m) and reset (r) resonators: Purcell filter frequency $\omega_{PF*}/2\pi$, resonator frequency $\omega_{*}/2\pi$, quality factor of the Purcell filter $\rm{Q}_{PF*}$ and the coupling rate of the resonator to Purcell filter $\rm{J}_{*}/2\pi$. 
We obtain the dispersive shift of readout and transfer resonators $\chi_{*}/2\pi$ by performing resonator spectroscopy with the qutrit initially prepared in the g, e and f state.
The coherence times of the qutrit are extracted from time resolved measurements.}
\label{tab:overviewSampleParameter}
\end{table}
The sample design is similar to the one used in Ref.~\onlinecite{Walter2017}. We etch the $\lambda/4$ coplanar waveguide resonators and feed-lines from a thin niobium film on a sapphire substrate using standard photolithography techniques. The transmon capacitor pads and Josephson junctions are fabricated using electron-beam lithography and shadow evaporation of aluminum. The parameters of the readout circuit (green elements in Fig.~\ref{fig:diagrams}a) and reset circuit (blue elements in Fig.~\ref{fig:diagrams}a) are obtained from fits to the transmission spectrum of the respective Purcell filter using the technique and model discussed in Ref.~\onlinecite{Walter2017} and are listed in Table~\ref{tab:overviewSampleParameter}. We extract the coupling strength of the transmon to both circuits using the same fitting procedure while preparing the transmon in its ground or excited state. The transition frequency $\omega_{\rm{ge}}/2\pi$, the anharmonicity $\alpha$ and the coherence times $T^{\rm{R}}_{\mathrm{2ge}}$, $T^{\rm{R}}_{\mathrm{2ef}}$ are measured using Ramsey-type measurements. The energy decay time $\rm{T}^{\rm{ge}}_{1}$ ($\rm{T}^{\rm{ef}}_{1}$) is extracted from an exponential fit to the measured time dependence of the populations when preparing the qubit in ether $\ket{e}$ or $\ket{f}$. The population $n_{\rm{th}}$ of state $\ket{e}$ in thermal equilibrium is extracted with the Rabi population measurement (RPM) method introduced in Ref.~\onlinecite{Geerlings2013}. We used a miniature superconducting coil to thread flux through the SQUID of the transmon to tune $\omega_{\rm{ge}}/2\pi$.
\section{Rabi Rate Extraction}
\label{app:fits}

In the fourth calibration step discussed in the main text, to measure the linear relation between the drive rate $\tilde{g}$ and drive amplitude $V_\mathrm{f0g1}$, we perform Rabi oscillation measurements (\cref{fig:resetCalib}~d and f).
To analyze these oscillations, we use a two-level model with loss described by the non-Hermitian Hamiltonian
	\begin{equation}
	H = 
	  \begin{bmatrix}
    i \gamma/2  & \tilde{g} \\
		\tilde{g}^*	 & i \kappa/2
		\end{bmatrix}
	\label{eq:f0g1Rabi}
	\end{equation}
which acts on states $\ket{f,0}$ and $\ket{g,1}$, analyzed in a rotating frame. The non-Hermitian terms $i \kappa/2$ and $i \gamma/2$ account for photon emission and transmon decay from $\ket{f}$ to $\ket{e}$, which bring the system to the dark states $\ket{g,0}$ and $\ket{e,0}$, respectively. Based on this model we derive an analytical expression for the $\ket{f}$ state population as a function of time
\begin{equation}
P_{\rm{f}}(t)=
\rm{e}^{-\frac{(\kappa+\gamma)}{2}t} \left|\cosh\left(\frac{\Omega t}{2}\right)+\frac{\kappa-\gamma}{2\Omega}\sinh\left(\frac{\Omega t}{2}\right)\right|^2
\label{eq:f0g1RabiPf}
\end{equation} 
where $\Omega=\sqrt{-(2\tilde{g})^2+(\kappa-\gamma)^2/4}$ is real positive or imaginary depending on the drive rate $\tilde{g}$. Using $P_{\rm{f}}(t)$ we obtain the fit function
\begin{equation}
f_{\rm{\tilde{g}}}(t)=\lambda P_{\rm{f}}(t-t_0) + \mu
\label{eq:f0g1fitfunction}
\end{equation} 
where the parameters $\lambda$ and $\mu$ account for potential state preparation and measurement (SPAM) errors and the parameter $t_0$ accounts for the fact that the gaussian rising and falling edges of the flat top f0-g1 pulse drive the f0-g1 transition for a finite time. 
For each drive amplitude $V_{\rm{f0g1}}$, we obtain Rabi oscillation data which we fit with \cref{eq:f0g1fitfunction}. To reduce the number of free parameters, we fit all data sets simultaneously and constrain $\lambda$, $\mu$, $t_0$ and $\kappa$ to be the the same for all sets, as these parameters are expected to be independent of $V_{\rm{f0g1}}$.

In the second calibration step discussed in the main text, we measure the linear dependence of the drive rate $\Omega_{\rm{ef}}$ on the drive amplitude $V_\mathrm{ef}$, by performing Rabi oscillation measurements (\cref{fig:resetCalib}~b and d).
We fit the time-dependence of the population $P_\mathrm{e}$ with the function
\begin{equation}
f_{\Omega_{\rm{ef}}}(t)=
\frac{1}{2}\rm{e}^{-\gamma_a t^{*}} \left(1-\rm{e}^{-\gamma_b t^{*}}\cos\left(\frac{\Omega_{\rm{ef}}t^{*}}{2}\right)\right).
\label{eq:efRabi}
\end{equation} 
where $t^{*} = t-t_0$ rescales the time $t$ by an offset $t_0$ to account for the fact that the rising and falling edges of the e-f pulse drive the e-f transition for a finite time. The parameters $\gamma_a$ and $\gamma_b$ account for transmon relaxation to $\ket{g}$ and decoherence in the $\{\ket{g},\ket{e}\}$ subspace, respectively. 
We verified numerically that \cref{eq:efRabi} is a good approximation of the time dependence of $P_\mathrm{e}$ during e-f Rabi oscillations and that it yields an unbiased estimate of $\Omega_{\rm{ef}}$, by comparing it to the result of a master equation simulation.
Similarly to the f0-g1 Rabi rate calibration, we simultaneously fit the Rabi oscillation data sets obtained for all probed $V_\mathrm{ef}$, constraining the fit parameter $t_0$, $\gamma_a$ and $\gamma_b$ to be the the same for all sets.

\section{Single-Shot Readout}
\label{app:readout}

\begin{figure}[t!]
\centering
\includegraphics{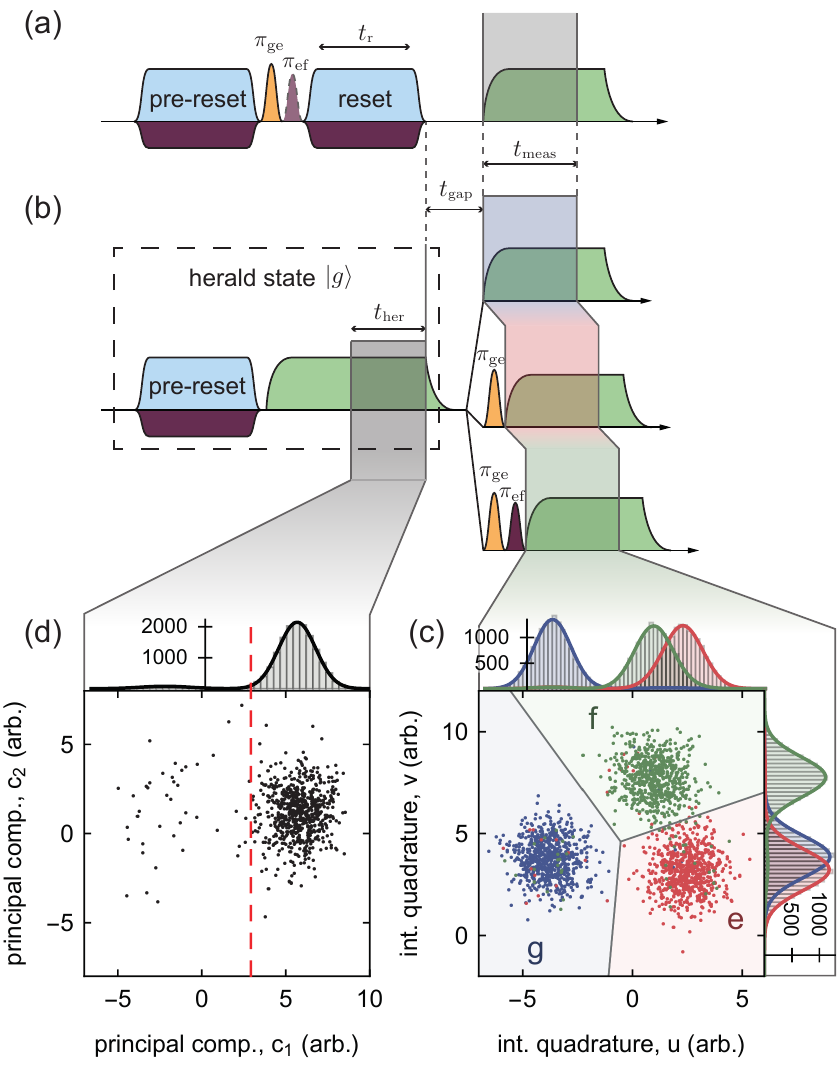}
\caption{
(a) Schematics of the pulse scheme used to test the unconditional reset protocol
(b) Schematics of the pulse scheme used to record reference single-shot counts.
(c) Subset of 500 reference traces displayed in the \(u\)-\(v\) plane, when the qutrit is prepared in state $\ket{g}$ (blue dots), $\ket{e}$ (red dots) or $\ket{f}$ (green dots). The assignment regions corresponding to label \(g\), \(e\) and \(f\) are shaded in blue, red and green respectively, and are separated by a gray line at their boundaries. 
(d) Sub-sample of 1000 traces acquired during the pre-selection pulse, projected in the principal component plane \(c_1\)-\(c_2\). Here \(c_1\) and \(c_2\) are the two first principal component of the set of traces. The red dashed line indicates the threshold for selection/rejection of the traces.
The sub-plots on the top or right axes of (c) and (d) show histogram counts of the traces. The solid lines in the sub-plots correspond to the density of the marginal probability distributions of the traces, scaled to match the histograms.
}
\label{fig:readout}
\end{figure}
To study the reset dynamics (\cref{fig:results}), we pre-reset the transmon with an unconditional reset, 
and prepare it in state $\ket{e,0}$ or $\ket{f,0}$ with a sequence of $\pi$-pulses (\cref{fig:readout}~a). Next, we apply the reset pulses for a duration $t_{\rm{r}}$, wait for a time $t_{\rm{gap}}$ and apply a microwave tone at the readout resonator to readout the transmon. We record the $I$ and $Q$ quadratures of the readout signal for a time $t_{\rm{m}}=120\,\rm{ns}$ starting at the rising edge of the readout tone. We refer to each recorded readout signal as a single-shot trace $S$.

To define an assignment rule that discriminates the transmon state based on a single-shot trace, we collect reference sets of 40000 single-shot traces obtained with the tranmon initialized in state $\ket{g}$, $\ket{e}$ and $\ket{f}$. State initialization is performed using a pre-selection readout pulse that heralds the transmon in its ground state (details discussed later in this section) followed by control $\pi$-pulses to prepare states $\ket{e}$ and $\ket{f}$ (\cref{fig:readout}~b). 
We integrate each reference single-shot trace with the weight functions $w_1$ and $w_2$, to calculate the integrated quadratures $u=\int_0^{t_{\rm{m}}} S(t) w_1(t) dt$ and $v=\int_0^{t_{\rm{m}}} S(t) w_2(t) dt$,  in post-processing. For each preparation state $\ket{p}$, the set of integrated traces $\vec{x}=(u,v)$ forms three clusters in the $u$-$v$ plane (\cref{fig:readout}~c) following a trimodal Gaussian distribution of mixture density
\begin{equation}
f_{\rm{p}}(\vec{x})=\sum_s{\frac{A_{\rm{s,p}}}{2\pi \sqrt{|\Sigma|}} e^{-\frac{1}{2}(\vec{x}-\mu_s)^\top\cdot\Sigma^{-1}\cdot(\vec{x}-\mu_s)}}
\label{eq:densityfunction}
\end{equation}
We extract the parameters $A_{\rm{s,p}}$, $\Sigma$ and $\mu_s$ with a maximum likelihood estimation. Based on these parameters, we define regions in the $u$-$v$ plane used to assign the result of the readout trace: if an integrated trace $\vec{x}_i$ is in the region labeled $m$, we assign it state $m$ (\cref{fig:readout}~c). 
By counting the number of traces assigned the value $m$ when the qutrit was prepared in state $\ket{s}$, we estimate the elements $R_\mathrm{m,s}=p(m|\ket{s})$ of the reference assignment probability matrix $R$  (see \cref{tab:readout}). 
\begin{table}[t]
\begin{centering}
\begin{tabular}{|c|ccc|}
\cline{1-4} 
\multicolumn{1}{|c}{} & \hfill{}$ \ket{g} $\hfill{} & \hfill{}$\ket{e} $\hfill{} & \hfill{}$\ket{f} $\hfill{} \tabularnewline
\cline{1-4}  
g & 98.2 & 2.5 &2.4 \tabularnewline
e & 0.9& 95.7 & 4.6 \tabularnewline
f & 0.9 & 1.8 & 93.0 \tabularnewline
\cline{1-4} 
\end{tabular}
\par\end{centering}
\caption{\label{tab:readout} Reference assignment probability matrix of identifying prepared states (columns) as the measured states (rows). The diagonal elements show correct identification, the off-diagonal elements misidentifications.}
\end{table}

To extract the the qutrit state populations $P=(P_\mathrm{g},P_\mathrm{e},P_\mathrm{f})$ after a SQUARE of duration $t_\mathrm{r}$, we also repeat the scheme illustrated in \cref{fig:readout}~a 40000 times, and record single-shot traces for each run. As for the reference sets, the assignment probability $M_\mathrm{m}$ is estimated by counting the number of traces assigned the value $m$ and follows
\begin{equation}
M_\mathrm{m}= p(m|P)  = \sum_s R_\mathrm{m,s}\cdot P_\mathrm{s}
\label{eq:readout}
\end{equation}
which can be expressed as $M=R\cdot P$. A simple approach to estimate the population $P$ of the qutrit is to set $P=M$. This approach is, however, sensitive to assignment errors due to readout imperfections: $P=M$ holds true only if $R_\mathrm{m,s}=\delta_\mathrm{m,s}$. To account for readout errors, we invert \cref{eq:readout} and set $P=R^{-1}\cdot M$.
However, this procedure relies on the accurate characterization of $R$, which is directly sensitive to errors in state-preparation for the reference trace sets. 
The qutrit therefore needs to be initialized in $\ket{g}$ before applying the reference readout tone, with a residual excitation that can be bounded, and that is ideally smaller than that of the unconditional reset protocol presented in this manuscript.
As mentioned earlier in this section, to do so, we pre-reset the transmon with our protocol
then herald the ground state of the transmon with a pre-selection readout pulse (\cref{fig:readout}b).
We record single-shot traces during the last $72\,\mathrm{ns}$ of the pre-selection pulse $t_{\rm{her}}$. The pre-selection traces form two clusters, corresponding to ground and excited traces, that are maximally separated along their first principal component axis (\cref{fig:readout}c). We model the distribution of the first principal component $c_1$ of the traces with a bimodal Gaussian distribution and extract its parameters with maximum-likelihood estimation. Based on this model, we calculate a threshold value $c_\mathrm{thr}$ such that $p(c_1>c_\mathrm{thr}|\mathrm{exc})=10^{-5}$. Selecting only traces with $c_1>c_\mathrm{thr}$ heralds the ground state of the transmon. On the set of selected traces, the residual excitation of the transmon at the rising edge of the reference readout tone is therefore dominated by transmon thermalization, which occurs at rate $k_{\uparrow}/2\pi=5\,\mathrm{kHz}$ in our sample. We use the same waiting time $t_\mathrm{gap}$ between initialization and readout to characterize the unconditional reset  
dynamics (\cref{fig:readout}a) and the reference trace set (\cref{fig:readout}b). As a result, thermalization occuring during this time can be seen as a source of readout error which is compensated. 
State preparation errors are then mostly explained by transmon thermalization occuring during the pre-selection, which we can bound by $k_{\uparrow} t_\mathrm{her}\simeq 0.25\%$. 

In conclusion, the corrected single-shot readout method we developed suffers from state preparation error resulting in a systematic under-estimation of the extracted populations, bounded by $0.25\%$. This residual error is small compared to the populations extracted during the unconditional reset 
for most measured points; this readout method is therefore suitable for the analysis conducted in this letter (\cref{fig:results}~c). However for configuration A, the unconditional reset 
leads to smaller excitation than the heralded reset, which explains why the extracted $P_\mathrm{exc}$ drops below zero at long reset times.

\section{Limitations of the Reset Protocol}
\label{app:limits}
The steady-state excited population $P_\mathrm{exc}^\mathrm{sat}$ that can be reached with the unconditional reset 
has three main limitations: the temperature of the reset resonator, thermalization of the transmon and residual driving of the g-e transition with the e-f drive. These limitations are quantitatively modeled in our master equation simulation, but they can also be discussed qualitatively to understand their effects on the performance.

In the level diagram of \cref{fig:diagrams}b, the black arrow labelled $\kappa$, connecting $\ket{g,1}$ to $\ket{g,0}$ represents the decay of the reset resonator. A finite temperature $T_{\rm{rr}}$ of the reset resonator can be accounted for by representing an arrow in the opposite direction with rate $\kappa \rm{exp}[-\hbar\omega_r/k_b T_{\rm{rr}}]$. If the unconditional reset
is dominated by this limitation, it can be shown that the temperature of the transmon reaches $T_{\rm{rr}} \omega_\mathrm{ge}/\omega_\mathrm{r}$ at steady-state.

The effective temperatures of superconducting qubits are typically higher than the base temperature of the dilution refrigerator $T_{\rm{BT}}$ which results in, a thermalization rate $k_{\uparrow}$ of the qubit being higher than expected from $T_{\rm{BT}}$. At equilibrium, thermalization competes against decay and the qubit has an equilibrium excited population $n_\mathrm{th}= k_{\uparrow} T_1$. Similarly, for the unconditional reset 
protocol, the competition between thermalization and reset rate yields the steady-state excitation population $P_\mathrm{th}^\mathrm{ss}\sim k_{\uparrow}/\Gamma$. 
At steady state, the probability of a transmon thermalization event (jump from $\ket{g}$ to $\ket{e}$) occuring between times $\tau$ and $\tau+d\tau$ is $(1-P_\mathrm{th}^\mathrm{ss})k_{\uparrow}d\tau \simeq k_{\uparrow}d\tau$. Immediately after such an event happens, the excited population follows $P_\mathrm{exc}^\mathrm{H} (\tau+t) = (1,0,0)^{\top}\cdot\rm{e}^{-i H t}\cdot(1,1,0)$ where $H$ is the non-Hermitian Hamiltonian~\ref{eq:ResetDynamics}. Integrating over all possible time windows for a thermalization jump to occur, we obtain that $P_\mathrm{th}^\mathrm{ss}=\int_0^{+\infty}P_\mathrm{exc}^\mathrm{H} (\tau) k_{\uparrow}d\tau$, which tends towards $k_{\uparrow}/\Gamma$ for large drive rates. Using this method, we calculate $P_\mathrm{th}^\mathrm{ss}=0.26\%$, $0.46\%$ and $0.34\%$ for configuration A, B and C, respectively. The good agreement of the calculated $P_\mathrm{th}^\mathrm{ss}$ with the measured and simulated values of $P_\mathrm{exc}^\mathrm{sat}$ for all parameter configurations further supports our interpretation that transmon thermalization is the dominant factor limiting the final population after reset.

Driving the e-f transition during unconditional reset 
broadens also the g-e transition. The e-f drive being detuned from the g-e transition by approximately the anharmonicity $\alpha$ of the transmon, it drives the g-e transition residualy which leads to e-f drive induced thermalization. This results in a steady-state excited population on the order of $\Omega^2_\mathrm{ef}/(\Omega^2_\mathrm{ef}+\alpha^2)\cdot \Omega_\mathrm{ef}/\Gamma$. It is then possible to go to lower excitation values at a lower rate, and a trade-off between speed and reset fidelity has to be set.

\section{Master Equation Simulation}
\label{app:MES}

To model the transmon qutrit reset process numerically, we start with the Hamiltonian of a transmon dispersively coupled to a high bandwidth resonator. We add the two drive-induced couplings required for the unconditional reset 
protocol, \textit{i.e.}~a Rabi drive between the $\ket{e},\ket{f}$ states of the transmon combined with an effective coupling $\tilde g$ between the $\ket{f,0}, \ket{g,1}$ states of the trasnmon-resonator system~\cite{Pechal2014,Zeytinoglu2015}. We represent the transmon  as an anharmonic oscillator with annihilation and creation operators $\hb$, $\hbd$~\cite{Koch2007} which we truncate at the second excited state $\ket{f}$ and denote the annihilation and creation operators of the reset resonator $\ha$ and $\had$, respectively. In a rotating frame at $\omega_r$ for the resonator and $\omega_{ge} + \alpha/2$ for the transmon, the transmon-resonator system is described by the Hamiltonian
\begin{equation}\label{eq:Ham}
\begin{aligned}
\hH/\hbar =&\; - \frac{\alpha}{2} \hbd \hb + \frac{\alpha}{2} \hbd \hbd \hb \hb + 2 \chi_r \had \ha \hbd \hb\\
&+ \frac{\tilde g}{\sqrt 2}( \hbd \hbd \ha + \had \hb \hb) + \frac{\Omega_{\rm{ef}}}{\sqrt 2} (\hb\, \expo{i \alpha t/2} + \hbd \expo{-i \alpha t/2}) ,
\end{aligned}
\end{equation}
where $\alpha$ is the transmon anharmonicity, $\chi_r$ the dispersive coupling strength between the trasnmon and the resonator, and $\Omega_{\rm{ef}}$ is the Rabi rate between the $\ket{e},\ket{f}$ states of the transmon. The readout resonator is omitted from the Hamiltonian since it does not affect the reset process and the induced static Lamb shifts are implicitly included in the parameters.

Numerical results are obtained by initializing the sytem in the $\ket{e,0}$ state and integrating the master equation 
\begin{equation}\label{eq:ME}
\begin{aligned}
\dot \rho/\hbar =& -i[\hH,\rho]\\
& + \kappa \mathcal D[\ha]\rho + \kappa_{\rm{int}}\mathcal D[\ha]\rho\\
& + \gamma_{1ge}(1 + n_{\rm{th}}) \mathcal D\left[\ket{g}\bra{e}\right]\rho + \gamma_{1ge} n_{\rm{th}} \mathcal D\left[\ket{e}\bra{g}\right]\rho\\
& + \gamma_{1ef}(1 + n_{\rm{th}})\mathcal D\left[\ket{e}\bra{f}\right]\rho + \gamma_{1ef} n_{\rm{th}} \mathcal D\left[\ket{f}\bra{e}\right]\rho\\
& + \gamma_{\phi \rm{ge}}\mathcal D\left[\ketbra{e} - \ketbra{g}\right]\rho \\
&+ \gamma_{\phi \rm{ef}}\mathcal D\left[\ketbra{f} - \ketbra{e}\right]\rho,
\end{aligned}
\end{equation}
where $\mathcal D[\hat O]\bullet = \hat O\bullet \hat O^\dag  - \{\hat O^\dag \hat O,\bullet\}/2$ denotes the dissipation super-operator, $\kappa_{\rm{int}}$ the internal decay rate of the resonator, $\gamma_{1nm}=1/T_{1nm}$ the decay rates of the transmon between the $\ket{n},\ket{m}$ states, $\gamma_{\phi \rm{nm}}=1/2T_1^{\rm{nm}}- 1/T_2^{\rm{nm}}$ the dephasing rates between the $\ket{n},\ket{m}$ states of the transmon and $n_{\rm{th}}$ the thermal population of the transmon qubit in steady state.

\newpage
\bibliography{Q:/USERS/Paul/04_Literature/RefDB/QudevRefDB}
\end{document}